\documentclass{article}

\usepackage{PRIMEarxiv}
\usepackage{microtype}      
\usepackage{hyperref}       
\usepackage{url}            
\usepackage[onehalfspacing]{setspace}

\usepackage{algorithm}
\usepackage{algpseudocode}
\usepackage{multirow}
\usepackage{tabularx} 
\usepackage{makecell}
\usepackage{graphicx}
\usepackage{booktabs}
\usepackage{svg}
\usepackage{amsmath}
\usepackage{txfonts}
\usepackage{ragged2e}
\usepackage[numbers]{natbib}
\usepackage[utf8]{inputenc} 
\usepackage[T1]{fontenc}    
\usepackage{hyperref}       
\usepackage{url}            
\usepackage{booktabs}       
\usepackage{amsfonts}       
\usepackage{nicefrac}       
\usepackage{microtype}      
\usepackage{graphicx}

\def\Authors{Shin Sano\,$^{1,3*}$ and Seiji Yamada\,$^{1,2}$}
\def\Address{$^{1}$Department of Informatics, School of Multidisciplinary Sciences,\\ Graduate University for Advanced Studies (SOKENDAI), Hayama , Japan \\
$^{2}$Digital Content and Media Sciences Research Division, National Institute of Informatics , Tokyo, Japan \\ $^{3}$Institute for Creative Integration , Oakland, California, USA}

\begin{document}

\renewcommand\theadalign{bc}
\renewcommand\theadfont{\bfseries}
\renewcommand\theadgape{\Gape[4pt]}
\renewcommand\cellgape{\Gape[4pt]}
\algnewcommand{\IIf}[1]{\State\algorithmicif\ #1\ \algorithmicthen}
\algnewcommand{\endIIf}{\unskip\ \algorithmicend\ \algorithmicif}
\algnewcommand{\FFor}[1]{\State\algorithmicfor\ #1\ \algorithmicdo}
\algnewcommand{\endFFor}{\unskip\ \algorithmicend\ \algorithmicfor}

\pagestyle{fancy}
\thispagestyle{empty}
\rhead{ \textit{ }} 
\fancyhead[LO]{Character Space Construction}  

\title{Composing Mood Board with User Feedback in Concept Space}

\author{
\Authors{}
\\
\\
\Address{} 
}
\maketitle


\begin{abstract}
We propose the Mood Board Composer (MBC), which supports concept designers in retrieving and composing images on a 2-D concept space to communicate design concepts. The MBC allows users to iterate adaptive image retrievals intuitively.  Our new contribution to the mood board tool is to adapt the query vector for the next iteration according to the  user's rearrangement of images on the 2-D space.  The algorithm emphasizes the meaning of the labels on the x- and y-axes by calculating the mean vector of the images on the mood board multiplied by the weights assigned to each cell of the $3 \time 3$ grid. The next image search is performed by obtaining the most similar words from the mean vector thus obtained and using them as a new query.  In addition to the algorithm described above, we conducted the participant experiment with two other interaction algorithms to compare. The first allows users to delete unwanted images and go on to the next searches.  The second utilizes the semantic labels on each image, on which users can provide negative feedback for query modification for the next searches.  Although we did not observe significant differences among the three proposed algorithms, our experiment with 420 cases of mood board creation confirmed the effectiveness of adaptive iterations by the Creativity Support Index (CSI) score. 
\end{abstract}


\section{Introduction}
Mood boards are visual artifacts often used as design development tools to communicate and share design ideas between stakeholders\cite{lucero2012framing}. Designers explore a wide range of visual images and select collections of images best-representing emotions, feelings, or ``moods" evoked by the original design brief or the brief as it develops. Mood boards usually consist of found and or made images for presentation. Abstract imagery is often more successful for this than figurative imagery, which can have strong literal interpretations\cite{garner2001problem}. Much prior work has explored processes and methods for assisting designers in creating visual mood boards that convey the intended design concepts. We will first introduce studies regarding mood boards in design education and industry practices.

Mood boards used to be created using analog and physical methods back in the days before graphic software became widely available. Mood boards are traditionally created by gluing together different types of media, including pictures from magazines, photographs, fabric, inspirational objects, etc. \cite{edwards2009comparative}

As \citet{cassidy2011mood} points out, mood boards are often used in the fashion industry and fashion-related or consumer product industries to communicate information visually. This is because mood boards play crucial roles in communicating more of the styling and aesthetic components of design rather than the functionality and practical components of design. 

\begin{figure*}[t]
 \centering
 \includegraphics[width=0.85\linewidth]{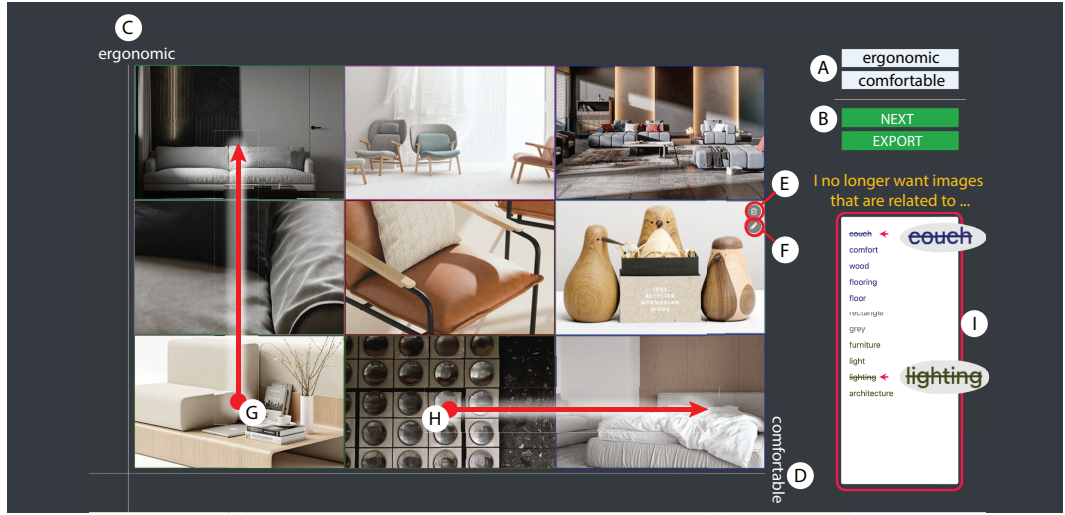}
 \caption{The MBC system UI with ``Reference 2'' algorithm, which involves the most elaborate user feedback. The proposed system does not have the label feedback feature(I). The proposed system allows users to move any image within the $3\times3$ matrix. Moving images upward (G) will enhance the semantics closer to the word $1$(C),  ``ergonomic'', and moving images to the right (H) will enhance the semantics closer to the word $2$ (D)``comfortable'' when it iterates the next search.}
\end{figure*}

Mood boards play multiple roles in design practice and education, such as thinking externalization, meaning acquisition, and conceptual reasoning \cite{li2021mood}. They are also used as qualitative design research tools facilitating creative thinking, presenting and communicating products \cite{cassidy2011mood}, expressing and communicating the designers' imagination and ideas they are pursuing \cite{edwards2009comparative}. Designers utilize them as a tool for communication with non-designers \cite{mcdonagh2004mood}. The processes of creating mood boards are more important to them than the results because they are intermediate mediums, not the final product of the design. \citet{bouchard2005nature} discuss the role of intermediate representations (IR) in design, where designers need to create multiple representations at different levels of abstraction.

\section{Related Work}
As in computational methods, several researchers have attempted to interactively support designers in composing visual mood boards with different focuses. A mood board-composing task involves a variety of algorithmic problems to solve, such as image retrieval, search strategies, computer vision, semantic feature engineering, natural language processing, and query expansion and modification based on user feedback. 

\citet{Setchi2011a} focused on the problem of the semantic gap, which is the discrepancy between the limited descriptive power of low-level image features and the richness of the semantics that users wish to operate\cite{smeulders2000content}. They proposed a semantic-based image retrieval approach that relies on textual information around the target image to avoid low-level and literal labels from given images. The method extracts the most relevant words in a document utilizing TF-IDF and uses a general-purpose ontology to expand the queries to find more of relevant images.

\citet{Koch2019a} focused on the problems in initial design ideation with an exploration-exploitation trade-off, which is the balance between adhering to the option that yielded the highest benefit in the past and exploring new options that might offer higher payoffs in the future\cite{MAL-024}. They created an interactive digital tool to support designers in creating a mood board, utilizing this exploration-exploitation strategy optimized by a cooperative contextual bandit reinforcement learning algorithm. The users can provide feedback to the suggested images by labeling them so that the system will take different query strategies based on the updated probability distribution of relevant suggestion agents for every iteration. 

\citet{Koch2020} further advanced the digital mood board tool and created the Semantic Collage, utilizing Google's Vision API to assign semantic labels to each image it suggests. It assists designers in translating ambiguous visual ideas into search terms and making sense of and communicating their designs. It allows designers to search images using verbal and visual queries and present semantic labels for each image. In addition, the user can give feedback to the system by modifying the weight of these labels, affecting the next search. One unique feature of the Semantic Collage is that the system presents labels for both individual images on the canvas and the collective semantics of all the images existing on the canvas, which allows designers to obtain hints on how to explain the mood board to the audience or discover new meanings through the board. 

The above two studies are remarkable in incorporating user feedback pragmatically based on closely observing user behaviors while engaged in the mood board creation task. Yet, no prior research has assumed a semantic space model on which designers can position their ideation relative to the verbal representation of a target design concept.

\begin{figure*}[ht]
 \centering
 \includegraphics[width=0.8\linewidth]{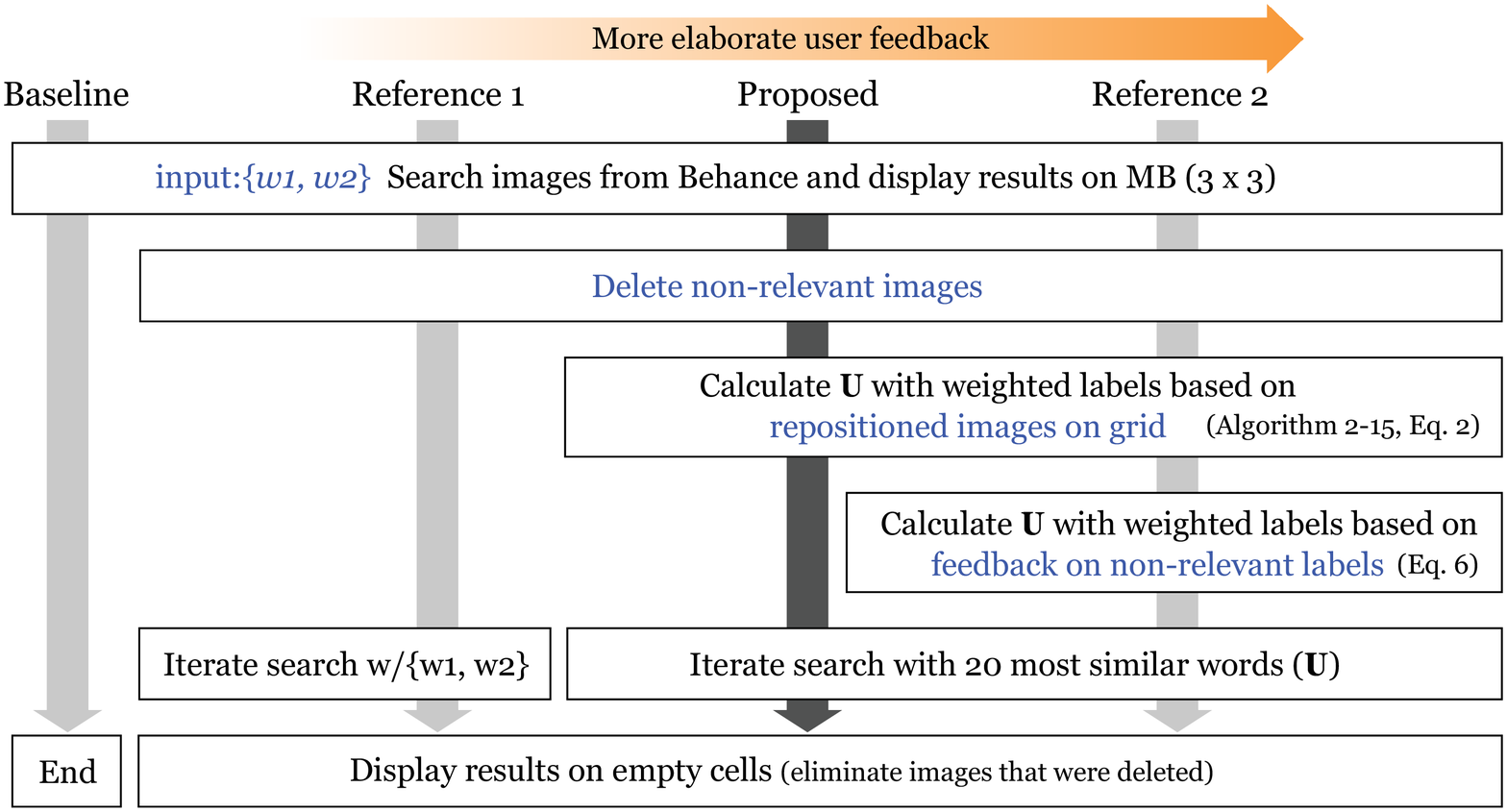}
 \caption{Variations of Mood Board Composer (MBC) algorithms. Text in blue shows the user's operations}
\end{figure*}

\section{Method}

The MBC is an AI-assisted interactive web application that helps designers compose a mood board derived from a Design Concept Phrase (DCP)\cite{sano2022ai}. Mood boards composed by MBC are constructed in a grid of $a\times b$ tiles. A few recent studies have indicated that participants in their studies typically handle 5 to 12 images per mood board\cite{Koch2020,Koch2019a,zabotto2019automatic}. \citet{aliakseyeu2006interacting} experimented with different sizes of digital image piles to compare human performances on navigation, repositioning, and reorganizing tasks and found significant differences in task performance between two different pile sizes (15 and 45). We chose nine images with a $3\times3$ grid to facilitate users in quick glancing and iterations while maintaining the capability to represent an original design concept with the images. The size of the mood board was also considered in terms of the  experiment logistics as we planned to conduct a large-scale participants' experiment. 

The MBC is designed to be used by concept designers who wish to explore and communicate their design concepts in visual representations. It also intends to build on the idea of the Character Space (CS) and the Design Concept Phrase (DCP) \cite{sano2022ai}, on which  users explore design concepts in a lexicosemantic space. The MBC assumes a DCP and searches images from Adobe Behance\cite{wilber2017bam}, a web-based portfolio service for professional and commercial artists. The UI renders the upper right quadrant of the Character Space (CS), consisting of word 1 and word 2 as attributions on the semantic axes. Therefore, the mood board images represent the design concepts expressed in the upper right quadrant of the CS (Fig.1-C,D). 

The proposed MBC system is designed to encourage users to iterate the exploration of images till they are satisfied with the overall mood board composition. Several studies have found that a design process involves internal thinking and external visualization\cite{sachse2009embodied} and iterative mental and manual processes\cite{rieuf2017emotional} with continuous improvement\cite{renaud2019framework}. Various cost factors can hinder these iterative processes, such as the time and effort to collect materials, trying different search queries and resources, and figuring out the compositions. \citet{edwards2009comparative} conducted a comparative study on developing physical and digital mood boards. They discussed that, even in the cases of digital mood board creation, the iterative process could be discouraged due to the vast choice of images offered by digital resources. They further argued that once images are selected, confidence is built so that users feel the continuous search for new material would be futile. This is caused by a type of confirmation bias, such that information is weighted more strongly when it appears early in a series\cite{baron2000thinking}. To confirm the positive effect of iterative processes and overcome this iteration cost, we designed our experiment in the following ways.

We first set up a comparative experiment between a baseline system that does not involve users' iterations for composing a mood board and the proposed systems, which allow users to iterate as often as they like. We aimed to implement low-cost and high-engagement interaction so that the users can effortlessly try the optimum number of iterations to get the best experience in mood board creation tasks. In addition, we developed versions of the MBCs as comparable references supporting different interactive feedback features to identify promising factors. The details of the features and the algorithms will be discussed in the following section.

\subsection{Overview of Systems}
The MBC tool consists of a front-end web application in JavaScript, HTML, and CSS and a back-end web server in Python hosted on the Amazon Elastic Compute Cloud (EC$2$)\cite{yi2010reducing}. Besides the baseline system, which does not support iterations, we developed the proposed algorithm and two reference algorithms implemented in the separate systems. Therefore, we have four variations for the tool factor. Figure 1 shows the overall differences in the procedures each variation takes. 

\subsection{Baseline Search Procedure}
Algorithm 1 and Fig. 2(Baseline) illustrate the procedure of the baseline search. It first receives the user's query($Q$) input as two adjectives, $(w_1, w_2)$, in the two search windows in the upper right section of the UI. When the ``START" button is pressed, the system will search images on Behance in three ``Fields,'' which are ``Industrial Design,'' ``Architecture,'' and ``Fashion.'' These three fields were selected because style elements such as form and CMF (Color, Material, and Finish) are important to the design of these fields. Many of the images in these fields visually represent those elements. Furthermore, other fields, such as web and graphic design, often contain textual information such as logos and copies. Images containing textual information that recalls a specific brand or image are unsuitable mood board material. The candidates of images($\mathit{images}$) are ranked by relevancy and sorted per field. The top nine images are then randomly assigned to an empty grid ($\mathit{grid}$) of the $3 \times 3$ image set ($D$) of the mood board. This single session concludes the procedure, and the user can export the mood board as a PNG file to a local client.

\begin{algorithm}
\caption{Mood Board Generation with Baseline Search}
\begin{algorithmic}[1]
\small
\Function{Baseline\_Search}{}
\State  $Q$ :=[], $D$ :=[], $\mathit{images}$ :=[] 
\State  $\mathit{grid}$:=\\$[(3,3),(2,3),(3,2),(2,2),(1,3),(3,1),(1,2),(2,1),(1,1)]$\\
\State  $Q$ := {\texttt{input}} $(w_1 + w_2)$
    
    \State \texttt{get}$\mathit{images}(Q)$from Behance
    \State \texttt{sort} Top\_$9 \mathit{images}$ by Relevancy
    \For{each $\mathit{grid}$ in $D$}    
    \State $\mathit{grid}$.append(Random($\mathit{images}$))
    \EndFor
    \State  \texttt{export}($D$)

\EndFunction
\end{algorithmic}
\end{algorithm}   

\subsection{Proposed algorithm - Query update with average vector calculation}
The proposed algorithm (Algorithm 2) involves query modifications based on user feedback. For each image on the current mood board, the system acquires semantic labels from the Google Vision API\cite{chen2017content}. The Vision API uses pre-trained machine learning models, assigns labels to images, and classifies them into millions of predefined categories. The proposed system obtains the top five labels for each image on the mood board, ranked by the confidence score.  Let $D\, (d\,_1, d\,_2, ..., d\,_m)$ be the image set on the current mood board, where $d\,_i$ is the $i$-th image on the mood board, $L^i\, (l\,^i_1, l\,^i_2, ...l\,^i_k)$ be the labels for each image, where $l\,^i_j$ is the $j$-th label for image $d\, ^i$, and $S\,^i\, (s\, ^i_1, s\, ^i_2,...,s\, ^i_k)$ be the confidence score from the Vision API assigned to each label, where $s\,^i_j$ is the score of the $j$-th label for image $d\,^i$. For each image label $l\,^i_j$ in the set of image labels $L^i$\, nested under each image $d\,_i$ on the mood board $D$, the system assigns label vectors using Concept Net Numberbatch word embedding. Let $\boldsymbol{V}\,^i\,(\boldsymbol{v}\,^i_1, \boldsymbol{v}\,^i_2,..., \boldsymbol{v}\,^i_k)$ be the vectors of the labels $L^i\, (l\,^i_1, l\,^i_2, ...l\,^i_k)$, where $\boldsymbol{v}\,^i_j$ is the vector of the $j$-th label for image $d\,^i$. A mean vector $\overline{\mathbf{v}}\,_i$ of the image $d\,_i$ can be calculated as follows: 

\begin{equation}
\begin{aligned}
\overline{\mathbf{v}}\,_i&=\frac{(s\,^i_1\,\boldsymbol{v}\,^i_1 + s\,^i_2\,\boldsymbol{v}\,^i_2 +,..., s\,^i_k\,\boldsymbol{v}\,^i_k)}{k}\\
&=\frac{1}{k}\sum_{j=1}^{k}\{s\,^i_j\,\boldsymbol{v}\,^i_j\}\\
\end{aligned}
\end{equation}
where $k$ is the total number of labels for image $d\,_i$.
\begin{algorithm}
\caption{Proposed (updating query)}
\begin{algorithmic}[1]
\small
\Function{NewQuery}{}

\State $\mathit{Q^{new}} :=[]$
\vspace{0.5mm}
\State $L, S, V :=[]$
\vspace{0.5mm}
\State $Wt := []$ 
\vspace{0.5mm}
\State $\overline{\mathbf{v}}_i :=[], \mathit{\scriptstyle{Weighted}}\,\overline{\mathbf{v}}_i :=[]$
\State $\mathbf{U} :=[]$\\

\For {each $d_i$ in $D$}
    \State $L$.append(VisionAPI($\mathit{d_i}$)) 
    \State $S$.append(VisionAPI($\mathit{d_i}$))
    \State $Wt$.append(OnDropWeight($x, y$))

        \For {each ${l_i}$ in $L$}
            \State $V$.append(ConceptNetVector($\mathit{l_i,  s_i}$))
            \If{$\mathit{cosSim}(l\,^i, w_1) > \mathit{cosSim} (l\,^i, w_2)$,\ }
            \State$\mathit{\scriptstyle{Weighted}}\,\overline{\mathbf{v}}_i = \overline{\mathbf{v}}_i \times Wt(\beta)$
                \Else
                \State $\mathit{\scriptstyle{Weighted}}\,\overline{\mathbf{v}}_i = \overline{\mathbf{v}}_i \times Wt(\alpha)$
            \EndIf
        \EndFor
    \State $\mathbf{U}$ := Mean($\mathit{\scriptstyle{Weighted}}\,\overline{\mathbf{v}}^i$)
\EndFor
\State $Q^{new}$.append(\textbf{MostSimilarWords}($\mathbf{U}$))
\EndFunction
\end{algorithmic}
\end{algorithm}   


The proposed systems let users reposition images on the mood board's $3 \times 3$ matrix. This operation determines which of the labels on images should be enhanced towards the semantics of either word 1 or word 2 by classifying the image labels into two classes, $w_1$\_labels, and $w_2$\_labels. Then, only one of the pairs of position weights, $Wt(\alpha, \beta)$ (Fig.3), assigned to each grid is multiplied for the labels that are classified as the class of label. This classification is performed by comparing the cosine similarity ($\mathit{CosSim}$) of each label to the vector of $w_1$ and $w_2$ (Algorithm 2-14). For example, suppose a label vector is more similar to the meaning of $w_1$. In this case, the label is classified as a $w_1$ label, and the label vector is multiplied only by the $\beta$ value ($w_1$ on $y$ axis side) of the pair of position weight $Wt(\alpha, \beta)$. This way, the user's repositioning an image towards a particular direction on the matrix will provide feedback to the system (Fig. 1). The system, in effect, will detect the users' intention to enhance a particular semantics in the following search without having to modify the query explicitly. The position-weighted average vector $\mathit{\scriptstyle{Weighted}}\,\overline{\boldsymbol{v}\,^i}$ of the repositioned image $d\,_i$ will be updated as described in Algorithm 2 (14-17). 
As for the paired weight for each position in the $3\times3$ grid, which will be multiplied by a label vector, we have tested two options with several initial queries. Figure 4 shows the weight array we implemented.  It keeps the images fairly close to the user's intention while expanding the semantic space to explore.

\begin{figure}[ht]    
 \centering
 \includegraphics[width=0.5\linewidth]{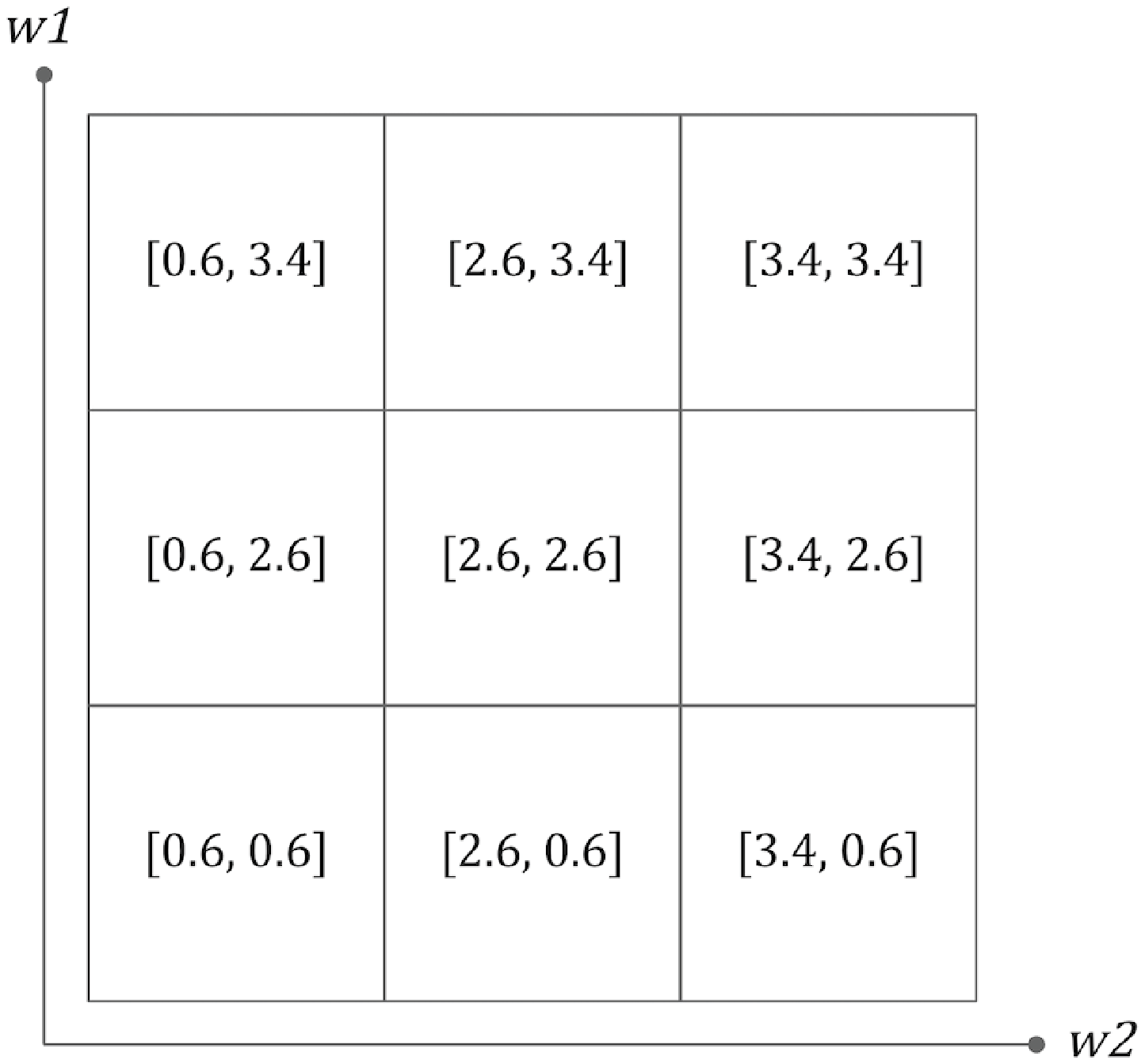}
\caption{Pairs of position weights $Wt(\alpha, \beta)$ on the mood board matrix. These weights are assigned upon dropping the image to $(x, y)$ coordinates.}
\end{figure}

The final step before updating the new query is to get the average vector of all the current images on the board, which can be calculated as follows. Let $\mathbf{U}$ be the average of all the weighted vectors for the images $\bigl\{\mathit{\scriptstyle{Weighted}}\,\overline{\mathbf{v}}_1, \mathit{\scriptstyle{Weighted}}\,\overline{\mathbf{v}}_2, ..., \mathit{\scriptstyle{Weighted}}\,\overline{\mathbf{v}}_m\bigr\}$ on the board. 

\begin{equation}
\mathbf{U} = \frac{1}{m}\sum_{i=1}^{m}\mathit{\scriptstyle{Weighted}}\,\overline{\mathbf{v}}\,_i
\end{equation}

where $m$ is the number of images on the current mood board.

\subsubsection{Calculating most similar words}
To update the query for the next search, the system will get the top $20$ most similar words according to the input, in this case, $\mathbf{U}$, the average vectors of all weighted vectors for images on the mood board $D$. The system computes the cosine similarity ($\mathit{CosSim}$) with the normalized input vectors and outputs the top-N words in $\mathit{CosSim}$. This function is implemented as a method in a Python package, gensim.models\cite{srinivasa2018natural}. This method computes $\mathit{CosSim}$ between a simple mean of the projection weight vectors of the given words and the vectors for each word in the model in the following procedure. 

Let $N$\ be the number of words in the pre-trained model, and $M$ be the dimensionality of the normalized vectors in the pre-trained model; then, the $N\times M$ matrix for words in the learned model is $X=(x_1, x_2, ..., x_N)$, and the input words can be expressed as $y$ in $M$-dimensionality. The $\mathit{CosSim}$ between $X_1$ and $y$ can then be calculated as follows.

\begin{equation}
\mathit{CosSim}(x_i, y) = \frac{x_i \cdot y}{\parallel x_i\parallel \cdot \parallel y \parallel}
\end{equation}

where $X_1$ is a normalized vector, so $\parallel x_i\parallel=1$. Therefore, the calculation of the $\mathit{CosSim}$ can be rewritten as:
\begin{equation}
\mathit{CosSim}(x_i, y) = \frac{x_i \cdot y}{\parallel y \parallel}
\end{equation}

Furthermore, calculating the $\mathit{CosSim}$ is more efficient if the matrix-vector product is used instead of element-by-element calculations. Therefore, the final $M$-dimensional $\mathit{CosSim}$ vector $\boldsymbol{S}$ is calculated using matrix-vector products as follows.

\begin{equation}
\boldsymbol{S} = \frac{X \cdot y}{\parallel y \parallel}
\end{equation}

\subsection{Reference 1 - no query modifications}
Reference 1 (Fig. 2) algorithm involves the least elaborate iteration procedure. Users can delete non-relevant images by clicking on the trash icon (Fig.1(E)) and try the next search by pressing the ``NEXT'' button (Fig.1(B)). The system will fill the empty cells with new images. This process can be repeated until the user is satisfied with the mood board. The function \textbf{NewQuery} is not implemented and keeps using the same initial query$Q (w_1 + w_2)$.

\subsection{Reference 2 - query modifications involving label feedback}

In addition to the proposed algorithm, the reference 2 version accepts user's feedback on semantic labels (Fig. 2).  When users click on the pencil icon on each image (Fig.1(F)), the semantic labels associated with each image are shown in the right bar (Fig.1(I)).  Users can rule out any non-relevant labels by crossing them out.  The method lets positive words contribute positively, and negative words contribute negatively towards the similarity computation. Let $n$ be the number of words that are crossed out as negative words in the labels, for example, in Fig. 1(I), ``couch'' and ``lighting.'' Since $m$ is the number of images with position-weighted average vectors on the current mood board, the new $\mathbf{U}$ is the mean vector of $m$ images and $n$ negative words together on the mood board. With the model Top-N most similar words, the new $\mathbf{U}$ can be calculated by multiplying $1$ by $m$ vectors and $-1$ by $n$ vectors, divided by the total number of the vectors, $m$ + $n$: 

\begin{equation}
\begin{aligned}
\mathbf {U} &=\frac{1}{m + n}{m\,\overline{\mathbf{v}}_i}*1+{\mathit{n}\,\mathit{v}\,^{negative}}*(-1)\\[3mm]
&=\frac{m\,\overline{\mathbf{v}}_i-n\,{\boldsymbol{v}\,^{negative}}}{m + n}\\
\end{aligned}
\end{equation}
\vspace*{0.5\baselineskip}

After scaling the mean vector $\mathbf{U}$ to the unit length ${\widehat{\mathbf{U}}}$, the system computes the dot product of $syn0norm$ (the vector collection of all words in ConceptNet Number Batch word embedding, normalized) to produce the Top-N, the most similar words to the mean vector.

\section{Experiment Design}
To investigate the MBC systems' effectiveness in supporting mood board composition tasks, we conducted a large-scale participant experiment using an online recruiting platform and a questionnaire platform. The following are the experiment specifications.

\subsection{Participants and Independent Variables}
$251$ participants, whose job function was ``Arts, Design, or Entertainment and Recreation'' and who was fluent in English, were recruited via Prolific. $32$ ($12.75\%$) did not complete the study due to system trouble or unknown reasons. This left us with a total of $219$ participants ($112$ M, $96$ F, $8$ Non-binary) who completed the study, with a mean age of $33.16$ years ($\sigma=11.07$). 
All of the participants who completed the study used either their own laptops or desktop computers. The participants who completed the study were paid US\$$12$. All participants were asked to perform the mood board creation task twice with the same type of MBC system; therefore, we collected 438 cases. The between-participant factor was the difference in the used MBC system (Fig.2), and the within-participant factor was the two different Design Concept Phrases (DCP) they were given to use as the initial query $Q$. The factor incorporated in these two DCPs was the $\mathit{CosSim}$ between word 1 and word 2 in the DCP. The near DCP was ``Ergonomic Comfortable,'' and the far DCP was ``Relaxed Skillful.'' The $\mathit{CosSim}$s of those two DCPs were $0.4528$ and $0.0053$, respectively. The order of the DCP they used in the two tasks was assigned randomly in a counterbalanced order. 

\subsection{Dependent Variables}
We used the Creativity Support Index (CSI)\cite{cherry2014quantifying} as a post-task psychometric measurement to compare four conditions, with a baseline MBC and three experiment MBCs, in terms of supporting creativity in a mood board composition task. The CSI enables us to quantify the cognitive processes of users using psychometric scales for six factors: Exploration, Expressiveness, Immersion, Enjoyment, Results Worth Effort, and Collaboration. The CSI evaluates a creativity support tool itself and focuses on the experience of using it to create rather than trying to evaluate a property of creative products directly.  Instead, the CSI evaluates the result of creation in relation to a user's effort, such as ``I was satisfied with what I got out of the system or tool."  This is suitable for tools designed for experienced users who know what the creative outcomes are and what the ideal experiences in creation are, as opposed to the tools designed for novice users where creativity is insufficient and thus needs support. The CSI also offers flexibility, such that it can be applied to various tools and scenarios over time and provides standardized measurement. It is robustly developed as each of the six factors has two different statements that improve the statistical power of a survey, which is deployed in the CSI. Combined with paired factor comparisons for each of the six factors, it provides insight into what aspects of creativity support may need attention. 

The CSI has a rigorous protocol that the researcher should follow for the measurement and analysis to be universal and reliable. For instance, our CSC tool is not designed for collaborative tasks; however, the CSI protocol discourages researchers from skipping the statements in the Collaboration factor, still allowing participants to rate the collaboration statements. Instead, the CSI protocol allows adding ``N/A'' responses to statements that belong to the collaboration factor, which we incorporated in our survey. 

In addition, we employed a single-item measurement for remaining mental resources, the Gas Tank Questionnaire (GTQ)\cite{monfort2018single}, immediately before and after each task. The GTQ attempts to measure users' cognitive load who engage in a task without burdening them by asking multiple questions. \citet{hart1988development} demonstrated that the NASA-TLX, a widely-used workload metric, increased the workload for participants who completed it and developed the GTQ as a low-burden alternative that does not affect the measurement itself negatively, especially when asked multiple times. The GTQ asks a question, ``Think about your brain as an engine. Slide the fuel tank indicator below to show how much gas you have left now.'' In our experiment, the GTQ questions were administered immediately before and after each mood board composition task, prompting the participant to slide the scale, which ranged from $0$ to $100$. We took the differences between Gas Tanks before and after the task as a value that indicated the mental resources consumed to perform the task. 

\subsection{Stimuli and Tasks}
Participants in all groups were recruited through Prolific and redirected to the Survey Monkey questionnaire platform. They were randomly assigned to either of the eight groups (two counter-balanced groups in different distance DCPs for each baseline, proposed, reference 1, and reference 2) types of tools  and given instructions on the experiment. After obtaining consent from each participant, the instructions were provided. The instructions were given in both videos embedded on YouTube and textual documents on Google Drive. The users were given options to use both or either of the instructions to ensure they would understand the nature of the tasks and how to use the MBC systems. 

The MBC tool was provided to the participant as a web link along with the DCP. The participants were asked to download the mood boards they created to their local computers and upload them to the questionnaire on Survey Monkey. They then went through all the CSI questionnaires. followed by the second pre-task GTQ, the second task with the same tool and the other DCP, and the second post-task GTQ. Finally, they responded to a Paired-Factor Comparison that gave weight to each category of CSI evaluations across both tasks.

\subsection{Participant Profile}
We designed the MBC with professional designers in mind. We were interested in examining CSI ratings regarding years of experience and how often the participants performed the mood board composition task we attempted to support with the MBC. For the former, we asked the participants how many years of experience they had in ``Arts, Design, or Entertainment and Recreation'' jobs. For the latter, we asked how often they perform mood board composition tasks as a categorical variable as follows: less than quarterly or never, quarterly, monthly, weekly, and daily.

\section{Results}
Note that of all the $438$ cases, $13$ cases were disqualified because their responses to the CSI questionnaire had identical scores all the way through the survey (all $0$ or all $10$), and two cases were excluded because they did not upload valid mood boards. Another three cases were omitted as outliers whose CSI scores were more than two standard deviations away from both sides of the mean, which left us with $420$ cases for the final analysis.

\begin{figure}[t]
    \centering
    \includegraphics[width=0.5\linewidth]{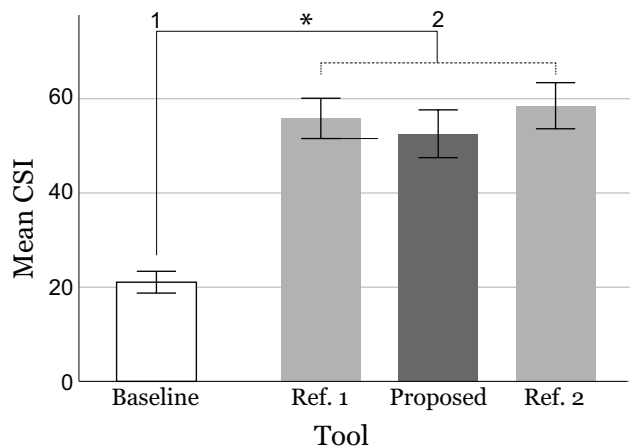}
    \caption{Mean CSI by the tool. The error bars show standard errors.}
    \label{fig:my_label}
\end{figure}

\begin{table}
    \centering
    \caption{Tukey HSD post-hoc comparison of individual groups.}
    \begin{tabular}{cccc} \toprule
        &\multicolumn{3}{c}{Mean difference $(p)$}\\
        \small{Tool}&\small{Proposed}&\small{Ref. 1}&\small{Ref. 2}\\
        \midrule
        \small{Baseline}&\makecell{$-30.66$\\$(<0.01{**})$}&\makecell{$-33.13$\\$(<0.01{**})$}&\makecell{$-36.05$\\$(<0.01{**})$}\\
        \hline
        \rule{0pt}{4ex} 
        \small{Proposed}&&\makecell[c]{$-2.46$\\$(0.861)$}&\makecell[c]{$-5.38$\\$(0.324)$}\\
        \hline        
        \rule{0pt}{4ex}         
        \small{Ref. 1}&&&\makecell[c]{$-2.92$\\$(0.778)$}\\
           
        \bottomrule
    \end{tabular}
\end{table}

\subsection{Creativity Support Index}
Fig. 4 shows the means of the CSI score by the tool. A one-way ANOVA demonstrated that the mean CSI score of the homogeneous subset of all the experiment groups: Proposed, Reference 1, and Reference 2, which were $55.83(\sigma=21.92)$, $52.57(\sigma=24.98)$, and $58.54(\sigma=24.95)$, respectively, were significantly higher (subset for $\alpha = 0.05$) than that of the baseline group ($21.05, \sigma = 12.36$). Table 1 shows post-hoc multiple comparisons by Tukey HSD. The mean differences show the CSI scores with the tools in the rows subtracted by that of the tools in the columns.

\begin{figure}[t]
 \centering
 \vspace*{0.5\baselineskip}
 \includegraphics[width={0.5\linewidth}]{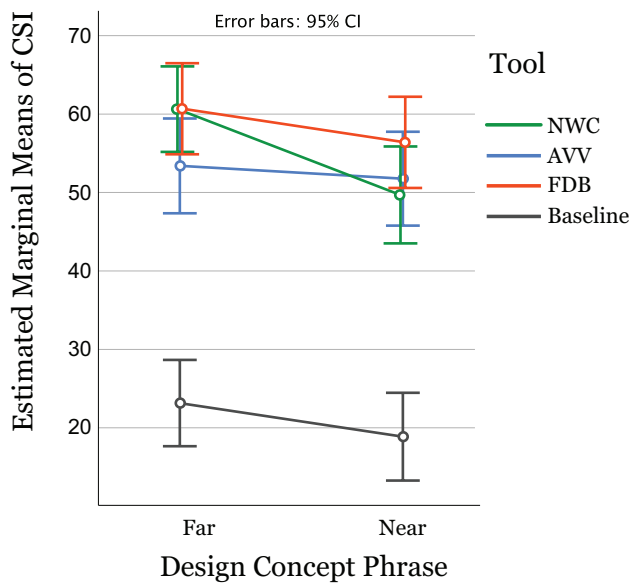}
 \caption{Variance in mean CSI for two different DCPs. Far DCP was ``Relaxed	Skillful''\;($\mathit{CosSim}=0.0029$), and near DCP was ``Ergonomic Comfortable''\;($\mathit{cosSim}=0.4528$)}
 \end{figure}

There were no significant differences in CSI scores among the proposed and reference tools. 


As for the difference in the number of iterations the participants performed by the tool, with the least elaborate tool, that is, Reference 1, participants engaged in iterations more significantly ($M=7.08, \sigma=5.08$) than with the subset of the proposed ($M=4.35,\sigma=4.04$) and reference 2 ($M=4.56,\sigma=3.90$). As for the mental resources they exhausted by performing the tasks, measured by the difference between the post-task GTQ and the pre-task GTQ, there was a significant difference ($p = 0.04$) between the baseline($-0.86,\sigma=9.9$) and the proposed tool($2.85,\sigma=8.74$). A difference was also observed tool between the baseline($-0.86,\sigma=9.9$) and the reference 2($2.57,\sigma=13.99$), with a significant tendency ($p = 0.065$).

\subsection{Within Participant Factor}
Fig. 5 shows the variance in the tool's estimated marginal means of the CSI scores for two different DCPs. Regardless of the tools in the Baseline or Experiment, the overall CSI score with the far DCP ($49.02,\sigma=25.04$) was significantly higher ($p=0.04$) than that with the near DCP ($43.41,\sigma=27.62$).


\section{Discussion}
The proposed algorithm, which allowed the participants to iterate the image search interactively, was valid in supporting creativity in the mood board composition task, demonstrated by the CSI score. Yet, we were interested in comparing the proposed algorithm with more and less elaborate feedback algorithms attempting to identify the most optimized formula, though there were no significant differences.
The values of the pre-task and post-task GTQ among the three experiment tools suggest that the users may have felt exhausted or overwhelmed by the complexity of the operation they had to follow on the MBC with the proposed and reference 2 algorithms. However, the weighted Result Worth Effort factor score by the tool shows that the subset of reference 2 ($M = 41.88,\sigma=22.48$) and reference 1 ($M = 40.47,\sigma=21.58$) had a significantly higher Result Worth Effort factor score than that of the proposed tool ($M = 32.79,\sigma=22.43$) alone (subset for $\alpha = 0.05$). This suggests that the users may have seen values in the process of reference 1 (deletion of non-relevant images) and reference 2 (label feedback) as more transparent.  On the other hand, the proposed system may have left users unclear about how repositioning images on the grid exactly works.  The reference 2 system involves labels attached to each image on the mood board, which is far more transparent about what the user is actually doing, crossing out labels so they will no longer see the images associated with the labels.

The matter of transparency and easier guidance to help users understand how the system works is crucial. In effect, how we explain a system to a participant before starting an experiment substantially affects the creative experience with digital tools. For the reference 2 algorithm, the participants were provided with the following explanation in text and video form: ``If you do not like a certain image, you can remove tags associated with the image from the search pool entirely. Click the pencil icon on the image, and strike out the labels you do not want to see. The system will avoid returning images containing those tags in future searches.'' However, considering our method, this is not an accurate explanation. We used word embedding vector calculation methods in Equation 6, where the $n$ vectors that receive negative feedback are multiplied by $-1$, and the $m$ average vector of each image on the current mood board is multiplied by $1$ and then divided by $m+n$. This is quite counterintuitive because label-level vectors are subtracted from image vectors. This method utilizes the principle of vector-space word representations, where algebraic operations can be performed on the word vectors, such that the vector ``King'' - vector(``Man'') + vector(``Woman'') results in a vector that is closest to the vector representation of the word ``Queen''\cite{mikolov2013efficient,mikolov2013linguistic}. However, explaining how this principle works accurately to participants could be intimidating and even discourage them from using it. Even a vague explanation like ``the feedback word will negatively affect the next search'' would not make sense. 

\begin{figure}[t]
 \centering
 \vspace*{0.5\baselineskip}
 \includegraphics[width={0.65\linewidth}]{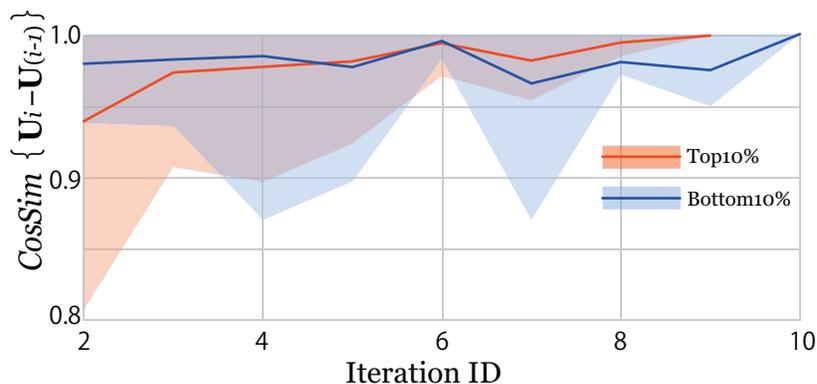}
 \caption{Range and mean cosine similarities ($\mathit{CosSim}$s) between iterations by CSI score. Cases with the top 10\% CSI tended to explore broader semantic space at the beginning and constantly converge as iterations progress. Cases with bottom 10\% CSI suggest the mean similarity between iterations stayed high, leading to less exploration, possibly resulting in unsatisfactory outcomes.}
 \end{figure}

The participants could have been impressed if the creativity support effect was substantial, even if some ambiguity and opaqueness were involved. Still, in this experiment, the effect was not strong enough to compensate for the increased task load and ambiguity indicated by the significant difference in GTQ value between the baseline and proposed tools.

As for the within-participants factor, we observed a significant advantage of the far DCP over the near DCP, regardless of the tool. This implies that a far DCP may yield a creative result by default that is more recognizable when users are engaged in a visual task. In fact, some of the factor scores in CSI showed significant differences; Result Worth Effort ($p = 0.037^*$): far DCP ($M = 36.98, \sigma = 21.26$), near DCP ($M = 32.42, \sigma = 23.12$); Enjoyment ($p = 0.042^*$): far DCP ($M = 26.48, \sigma = 22.72$), near DCP ($M = 22.00, \sigma = 22.29$). Note that Expressiveness also showed a difference tendency ($p = 0.055$).


 
In addition, we retrieved and analyzed the log data from the participant experiment to gain insight from the high-performers and low-performers on the mood board composition task. For the MBC tools with the proposed reference 1, and reference 2 algorithms, we implemented a system log that records the following data on each iteration: the $w_1$ and $w_2$, iteration ID, except for the initial search with the ($w_1$, $w_2$), image labels: $L^i\, $, confidence score: $S\,^i\,$, $\mathit{CosSim}\{w_1, \overline{\mathbf{v}}\,_i(a, b)\}$, $\mathit{CosSim}\{w_2, \overline{\mathbf{v}}\,_i(a, b)\}$, $\mathit{CosSim}(w_1,\mathbf{U})$, $\mathit{CosSim}(w_2,\mathbf{U})$, time stamp, $Top-N\_words\, \mathit{for}\, \mathbf{U}$, and negative feedback words. 

Fig. 6 shows the transition in the cosine similarity ($\mathit{CosSim}$) between the average vectors $\mathbf{U}_i - \mathbf{U}_{(i-1)}$ of the mood board, where $\mathbf{U}_h$ is the average mood board vector in the $i$-th iteration. The red lines represent the transition in $\mathit{CosSim}$s over the multiple iterations for the cases with the top $10\%$ CSI scores, which indicate that the change in the semantics of the mood board between iterations was greater at the beginning and became smaller towards the end of the iterations. This suggests that the participants in these cases changed more images at the beginning and fewer images towards the end to make more minor adjustments to the finish. On the other hand, the blue lines represent the transition in $\mathit{CosSim}$s over the multiple iterations in cases with the bottom $10\%$ CSI scores, indicating that the mood board's semantics did not change much from the beginning through the end. This implies that the participants in these cases kept changing a smaller part of the mood board and did not really reach a satisfactory set of nine images toward the end. The mean count of iterations for the top $10\%$ participants ($4.11,\sigma=2.08$) was significantly lower ($p = 0.05^*$) than that for the bottom $10\%$ participants ($6.69,\sigma=4.70$). The mean $\mathit{CosSim}$ between iteration $2$ and iteration $1$ for the top $10\%$ participants ($0.940,\sigma=0.0497$) was significantly smaller ($p < 0.01^{**}$), meaning that there was more distance than that for the bottom $10\%$ participants ($0.979,\sigma=0.0234$).

\section{Conclusion}
Through experimenting with the MBC tools with different levels of interactivity, we confirmed the effectiveness of user feedback, making the mood board creation task more engaging for concept designers.  We have contributed to the field of computational creativity tools by offering adaptive query updates utilizing the 2-D semantic space where users can rearrange the images on the mood board.  The variation of the tool that incorporated the user feedback on semantic labels was also valid. Our post-hoc analysis of factor scores of the CSI and GTQ questionnaire suggests the advantage of transparency in the relationship between action and outcomes from the users' perspective, despite the increased workload.  We also observed that the characteristics of the initial verbal query are a strong factor for users to feel creative about the concepts they are operating.  The analysis of the top and bottom performers revealed the difference in the process of semantic convergence across the iterations.  High performers seemed to tend to explore earlier and exploit later.

\bibliographystyle{unsrtnat}
\bibliography{Sano_bib_2.1.2023}

\end{document}